\documentclass[12pt,preprint]{aastex}

\def\vec#1{\mbox{\boldmath $#1$}}
\slugcomment{Not to appear in Nonlearned J., 45.}

\shorttitle{MHD Simulation of X2.2 solar flare on 2011 Feb./ 15}
\shortauthors{Inoue et al.}
 
\begin{document}


\title{Magnetohydrodynamic Simulation of the X2.2 Solar Flare \\ 
       on 2011 February 15: I. Comparison with the Observations}


\author{S.\  Inoue}
\affil{ School of Space Research, Kyung Hee University, Yongin 446-701, Korea}
\email{inosato@khu.ac.kr}

\author{K.\ Hayashi,}
\affil{W.\ W.\ Hansen Experimental Physics Laboratory, Stanford University \\
       Stanford, CA 94305 USA}

\author{T.\  Magara}
\affil{ School of Space Research, Kyung Hee University, Yongin 446-701, Korea}

\author{G.\  S.\  Choe}
\affil{ School of Space Research, Kyung Hee University, Yongin 446-701, Korea}

\and 

\author{Y.\  D.\  Park}
\affil{ Korea Astronomy and Space Science Institute, Daejeon 305-348, Korea}

% ========================================================================
% Abstract
% ========================================================================
  \begin{abstract}
  We performed a magnetohydrodynamic (MHD) simulation using a nonlinear 
  force-free field (NLFFF) in solar active region 11158 to clarify the 
  dynamics of an X2.2-class solar flare. We found that the NLFFF never shows 
  the drastic dynamics seen in observations, {\it i.e.,} it is in stable state 
  against the  perturbations. On the other hand, the  MHD simulation shows 
  that when the strongly twisted lines are formed at close to the neutral line,
  which are produced via tether-cutting reconnection in the twisted lines of 
  the NLFFF, consequently they erupt away from the solar surface via the 
  complicated reconnection. This result supports the argument that the 
  strongly twisted lines formed in NLFFF via tether-cutting reconnection are 
  responsible for breaking the force balance condition of the magnetic fields 
  in the lower solar corona. In addition to this the dynamical evolution of 
  these field lines reveals that at the initial stage the spatial pattern of 
  the footpoints caused by the reconnection of the twisted lines appropriately
  maps the distribution of the observed two-ribbon flares. Interestingly, 
  after the flare the reconnected field lines convert into the structure like 
  the post flare loops, which is analogous to EUV image taken by SDO. 
  Eventually, we found that the twisted lines exceed a critical height at 
  which the flux tube becomes unstable to the torus instability. These results
  illustrate the reliability of our simulation and also provide an important 
  relationship between flare-CME dynamics.
  \end{abstract}

% =============================================================================
% Introduction
% =============================================================================
  \section{Introduction}
   Solar active region (AR) 11158 emerged in the eastern solar hemisphere on 
  February 11, which is consists of bipolar fields showing the complicated 
  behaviors including the strongly sheared and twisted motions
  (\citealt{2012ApJ...744...50J}). Eventually, it produced one X-class flare 
  and several M-class flares on February 2011 
  (\citealt{2012ApJ...748...77S};
   \citealt{2013ApJ...772..115T};
   \citealt{2012ApJ...748L...6N};
   \citealt{2013ApJ...770...79I} and 
   \citealt{2014arxiv}). 
  Fortunately, solar physics satellites, the Solar Dynamics Observatory 
  ({\it SDO}), and {\it Hinode} (\citealt{2007SoPh..243....3K}) provide rich 
  data with unprecedented temporal and spatial resolutions obtained by 
  multi-wavelength observations, enabling us to further understand the 
  dynamics of these solar flares, {\it e.g.,}
   \cite{2011ApJ...738..167S},
   \cite{2011ApJ...734L..15K},
   \cite{2012ApJ...745L..18A},
   \cite{2012ApJ...745L...4L},
   \cite{2012ApJ...752L...9J},
   \cite{2013ApJ...770...79I} and 
   \cite{2013ApJ...773..128T}. 

   In general the solar flares are widely considered as release phenomena of 
  free magnetic energy accumulated in the solar corona
  (\citealt{2011LRSP....8....6S}), it's therefore important matter to 
  understand the three-dimensional (3D) magnetic field to know where the free 
  energy is accumulated in the AR. Unfortunately we cannot directly 
  measure the coronal magnetic field even by the state-of-art ground or space 
  observations which provide only 2D vector magnetic field on the photosphere. 
  In this kind of situation, nonlinear force-free field (NLFFF) extrapolation 
  from the vector field is a solid tool due to allow us to show 3D view of 
  the magnetic field (see \citealt{2012LRSP....9....5W} in details). The 
  Helioseismic and Magnetic Imager (HMI; \citealt{2012SoPh..275..207S};
  \citealt{2012SoPh..275..229S}; \citealt{2014SoPh_JTH}) on board a 
  ({\it SDO}) and the Solar Optical Telescope 
  (SOT; \citealt{2008SoPh..249..167T}) on board {\it Hinode} provide the 
  vector field data, which were already applied to extrapolate the 3D magnetic
  field under the NLFFF approximation by earlier studies 
  ( \citealt{2012SoPh..281...37W}; 
    \citealt{2013ApJ...769..144J}; 
    \citealt{2013ApJ...770...79I} and 
    \citealt{2014arxiv}).
  In common of them, they show  that the strongly twisted and sheared field 
  lines are found to reside at close to the polarity inversion line (PIL) 
  before the flare, which are confirmed to relax after the flare. 
      
   Some earlier studies analyzed the NLFFF and suggested the dynamics of the 
  M6.6 or X2.2-class flares due to compare the NLFFF  with multi-wavelength 
  observations or NLFFF configurations before and after the flares. 
  \citealt{2012ApJ...745L..17W} suggested the possibility of tether-cutting 
  reconnection (\citealt{2001ApJ...552..833M}) in X2.2-class solar flare 
  through multi-wavelength analysis. \cite{2012ApJ...748...77S} also analyzed 
  the NLFFF, showing the energy distribution of it in space and temporal 
  evolution as well as suggesting an evidence of the tether-cutting 
  reconnection on the variation of the field lines connectivity  before and 
  after the flare. \citealt{2012ApJ...745L...4L} analyzed the NLFFF just 
  before and after the M6.6-class solar flare and found that the strong 
  coronal current system underwent an apparent downward collapse and a small 
  current-carrying loop was produced closer to the surface, suggesting that 
  the tether-cutting reconnection is a suitable model. More recently, 
  \cite{2013ApJ...778L..36L} estimated the twist of a field line and 
  investigated the connectivity of a field line, consequently the obtained 
  results were consistent with the tether-cutting magnetic reconnection model. 

  \cite{2013ApJ...770...79I} also estimated the magnetic twist of the NLFFF 
  before the X2.2-class flare in AR 11158, reconstructed by MHD relaxation 
  method (\citealt{2014ApJ...780..101I}) and found that the strongly twisted 
  lines having more than a half-turn twist but less than one-turn are related 
  to this flare because their footpoints well correspond to the location 
  where the two-ribbon flares in {\ion{Ca}{2}} image taken by SOT/{\it Hinode} 
  are enhanced. This result indicates that the NLFFF is found to be stable 
  state against an ideal MHD instability, in other words, the tether-cutting 
  reconnection would be needed to generate the more strongly twisted lines 
  with more than one-turn twist, which might have a potential to produce the 
  eruptive phenomena.
     
   However, because these NLFFFs, which are constructed in the steady sate,  
  cannot reveal the dynamics in the solar flares, we cannot verify the 
  occurrence of tether-cutting reconnection. In this study, we explore the 
  MHD dynamics of the X2.2 solar flare on 2011 February 15, using MHD 
  simulation involving the NLFFF. This simulation can tell us the dynamics
  of the magnetic field in close to the real situation, and then enable us 
  to directly compare the results to observations. In this paper we show 
  the results focusing on an overview of the 3D dynamics of the magnetic 
  field and comparison of them with observations. The detailed dynamics, 
  {\it e.g.}, it associated with the complicated reconnection process during 
  flare will be shown in next paper. This paper is constructed as follows. 
  Our observational data and numerical method are described in Section 2. 
  Our results are presented in Sections 3 and discussed in Section 4. Our 
  conclusion is summarized in  Section 4. 
          
% ========================================================================
% Numerical method
% ========================================================================
  \section{Observations and Numerical Method}
  \subsection{observations}
  To extrapolate the NLFFF, we use the vector magnetic field shown in left 
  panel of Figure \ref{f1}(a), observed at 00:00 UT on 2011 February 15, 
  approximately 120 minutes before the X2.2-class flare detected by 
  HMI/{\it SDO}. The vector magnetic field covers a 216 $\times$ 216 (Mm$^2$) 
  region, divided by a 600 $\times$ 600 grid which is provided by the 
  description of CEA projected and remapped vector magnetic field
  \footnote{http://jsoc.stanford.edu/jsocwiki/ReleaseNotes}. It is obtained 
  using the very fast inversion of the Stokes vector (VFISV) algorithm 
  (\citealt{2011SoPh..273..267B}) based on the Milne$-$Eddington 
  approximation. A minimum energy method (
  \citealt{1994SoPh..155..235M}; 
  \citealt{2006SoPh..237..267M};
  \citealt{2009SoPh..260...83L}) was used to resolve 180$^\circ $ ambiguity in 
  the azimuth angle of the magnetic field. We also use extreme ultraviolet 
  (EUV) images observed with 94{\AA} and 171 {\AA} taken by Atmospheric 
  Imaging Assembly (AIA; \citealt{2012SoPh..275...17L}) on board {\it SDO} and
  {\ion{Ca}{2}} H 3968 {\AA} images during the flare taken by {\it Hinode} 
  , in order to compare with the NLFFF and MHD simulation. {\ion{Ca}{2}}
  images were obtained using the Broadband Filter Imager of the SOT on board 
  {\it Hinode}. The field of view corresponds to 111.58" $\times$ 111.58" with
  a resolution of 1024 $\times$ 1024 at 01:50:18 UT on February 15.
 
  \subsection {Numerical Method of the Nonlinear Force-Free Extrapolation}   
   We carry out the NLFFF and MHD simulation according to the following 
  equations,

% ------------------------------------------------------------------------
% Density
% ------------------------------------------------------------------------
  \begin{equation}
  \rho = |\vec{B}|
  \label{eq_den_NLFFF} 
  \end{equation}

or

  \begin{equation}
  \frac{\partial \rho}{\partial t} = -\vec{\nabla} \cdot (\rho \vec{v})
                                     + \xi \nabla^2 (\rho - \rho_0)
  \label{eq_den_MHD}
  \end{equation}

% ------------------------------------------------------------------------  
% Equation of Motion      
% ------------------------------------------------------------------------   
  \begin{equation}
  \frac{\partial \vec{v}}{\partial t}
                        = - (\vec{v}\cdot\vec{\nabla})\vec{v}
                          + \frac{1}{\rho} \vec{J}\times\vec{B}
                          + \nu\vec{\nabla}^{2}\vec{v}.
  \label{eq_motion}
  \end{equation}

% ------------------------------------------------------------------------ 
% Induction equation                    
% ------------------------------------------------------------------------   
  \begin{equation}
  \frac{\partial \vec{B}}{\partial t}
                        =  \vec{\nabla}\times(\vec{v}\times\vec{B}
                        -  \eta_{i}\vec{J})
                        -  \vec{\nabla}\phi,
  \label{induc_eq}
  \end{equation}

% ------------------------------------------------------------------------ 
% Ampere's low                                                 
% ------------------------------------------------------------------------ 
  \begin{equation}
  \vec{J} = \vec{\nabla}\times\vec{B}.
  \end{equation}

% ------------------------------------------------------------------------ 
% Dedner                                          
% ------------------------------------------------------------------------  
  \begin{equation}
  \frac{\partial \phi}{\partial t} + c^2_{h}\vec{\nabla}\cdot\vec{B}
    = -\frac{c^2_{h}}{c^2_{p}}\phi,
  \label{div_eq}
  \end{equation}
  where  NLFFF calculation employs the equation (\ref{eq_den_NLFFF})
  as the density evolution in order to ease the relaxation by equalizing 
  the Alfven speed in space as shown in  \cite{2013ApJ...770...79I}. 
  $\rho$ is mass density,  $\rho_0$ is an initial value of density 
  where this density evolution is  following \cite{2010ApJ...708..314A},   
  $\vec{B}$ is the magnetic flux density,  $\vec{v}$ is the velocity, 
  $\vec{J}$ is the electric current density, and $\phi$ is a scalar potential.
  The last equation (\ref{div_eq}) is used  to reduce deviation from 
  $\nabla\cdot\vec{B}=0$ and was introduced by 
  \cite{2002JCoPh.175..645D}. 

  The length, magnetic field, density, velocity, time, and electric 
  current density are normalized by   
  $L^{*}$ = 216   Mm,  
  $B^{*}$ = 2500 G, 
  $\rho^{*}$ = $|B_0|$,
  $V_{A}^{*}\equiv B^{*}/(\mu_{0}\rho^{*})^{1/2}$,    
  where $\mu_0$ is the magnetic permeability,
  $\tau_{A}^{*}\equiv L^{*}/V_{A}^{*}$, and     
  $J^{*}=B^{*}/\mu_{0} L^{*}$,  
  respectively. The non-dimensional viscosity $\nu$ is set as a constant  
  $(1.0\times 10^{-3})$, and the coefficients $c_h^2$, $c_p^2$ in 
  equation (\ref{div_eq}) also fix the constant values, 0.04 and 0.1, 
  respectively. Subscript $i$ of $\eta$ corresponds  to 'NLFFF' or 'MHD'.  
  These equation sets, normalize values and parameters are mostly identical 
  to the previous our studies \cite{2013ApJ...770...79I} except the density 
  evolution described in equation (\ref{eq_den_MHD}).
  
  A formulation of 
  $\eta_{NLFFF}$ is given by  
  \begin{equation}
  \eta_{NLFFF} = \eta_0 + \eta_1 \frac 
                {|\vec{J}\times\vec{B}||\vec{v}|^2}{\vec{|B|^2}},
  \label{eta_nlfff}
  \end{equation}
  which is slightly different from \citealt{2013ApJ...770...79I}, where 
  $\eta_0=1.0\times 10^{-5}$ and $\eta_1=1.0\times 10^{-3}$. The vector field 
  shown in Figure \ref{f1}(a) is preprocessed in accordance with 
  \cite{2006SoPh..233..215W}  and a potential field applied to an initial 
  condition of the NLFFF calculation is extrapolated using Green function 
  method (\citealt{1982SoPh...76..301S}). The observed tangential components 
  of the magnetic field reside in the region enclosed by a dashed square, as 
  shown the left panel in Figure \ref{f1}(a); the outside region is fixed by 
  the potential field, because we focused on the dynamics of the magnetic 
  field in an early phase near the central area and constructing a field 
  close to an equilibrium state numerically. Note that, according to the 
  mathematical formula,  this state mixing of the force-free field and the 
  potential field cannot keep an equilibrium. Nevertheless, we employed 
  this method in this study in order to avoid an effect from inconsistent 
  force-free alpha residing outside area where the weak magnetic fields are 
  dominated.
      
  \subsection{Numerical Method of the MHD Simulation}
   Next, we set MHD equations under the zero $ \beta$ approximation 
  (\citealt{1989ApJ...338.1148M} or \citealt{1996ApJ...466L..39A}) due to the 
  low $\beta$(0.01 $\sim$ 0.1) coronal plasma where the time evolution of 
  density is according to the equation (\ref{eq_den_NLFFF}) being similar with 
  \cite{1996ApJ...466L..39A}, {\it i.e.}; these equations are identical to the 
  NLFFF extrapolation method in \cite{2014ApJ...780..101I} or 
  \cite{2013ApJ...770...79I} except the velocity limit is not imposed and the 
  resistivity formulation is different. 

  Initial condition on the density is given by $\rho_0=\vec{|B|}$ in all MHD 
  (also NLFFF) calculations. In this case, the density is modeled by equation 
  (\ref{eq_den_NLFFF}); therefore, we cannot compare it with observational 
  data exactly or discuss the time scale of the dynamics of the magnetic 
  field, as the Alfven time scale depends on this model. However, this formula
  might be more reasonable than that in the constant density, {\it e.g}, 
  $\rho=$1.0 because the real coronal density is stratified due to gravity, 
  then Alfven wave is propagation effectively in the upper corona where the 
  magnetic field is driven there. In addition to these, 
  \cite{2006ApJ...645..742I} showed the dynamics of a flux tube using two 
  types of density; one is modeled by $\rho=1.0$ at every time,  {\it i.e.}, 
  the density variation is not considered. Another is obtained from a 
  continuous equation (\ref{eq_den_MHD}). Consequently, both dynamics of the 
  magnetic field are nearly identical at the early eruptive phase. We will 
  discuss later again in this paper. A resistivity formula is given as an 
  anomalous resistivity as following,    
  \begin{equation}
  \eta_{MHD}  = \left\{
         \begin{array}{ll}
         \eta_{0} & \quad \mbox{$J<j_{c}$}, \\
         \eta_{0}+\eta_{2} 
              (\frac{J-j_{c}}{j_{c}})^{2} & \quad \mbox{$J>j_{c}$},
         \end{array}\right.
   \label{ano_res}
   \end{equation}
   where $\eta_{2}=5.0\times 10^{-4}$, and $j_{c}$ is the threshold current,
    set to 30 in this study. 
  
   In this simulation, a numerical box with dimensions of 216 $\times$ 216 
   $\times$ 216 (Mm$^3$) is given by 1 $\times$ 1 $\times$ 1 as a 
   non-dimensional value. The grid number is assigned as 300 $\times$ 300 
   $\times$ 300, which is 2 $ \times$ 2 binning from the original data. 
   Regarding to the boundary conditions of NLFFF (Run A) and MHD relaxation 
   processes (Run C), physical values on all of the boundaries 
   are fixed (fixed boundary condition), except for the Neumann-type boundary 
   condition on which the convenient potential $\phi$ is imposed. On the other 
   hand, in MHD simulations (Run B, and Run D-F), tangential components of the 
   magnetic field are released on all of the boundaries (Released boundary 
   conditions). The numerical scheme is quite identical to that in 
   \cite{2014ApJ...780..101I} or \cite{2013ApJ...770...79I}. The kind of 
   simulation (NLFFF or MHD), equation of the density evolution,  initial 
   conditions, and boundary conditions used in each Run  are summarized in 
   table {\ref{tbl-1}}.

% =============================================================================
% Results
% =============================================================================
  \section{Results}

% -----------------------------------------------------------------------------
  \subsection{Results of the Nonlinear Force-Free Extrapolation}
% -----------------------------------------------------------------------------
  \subsubsection{3D Magnetic Structure of NLFFF}
\begin{figure}
  \epsscale{.95}
  \plotone{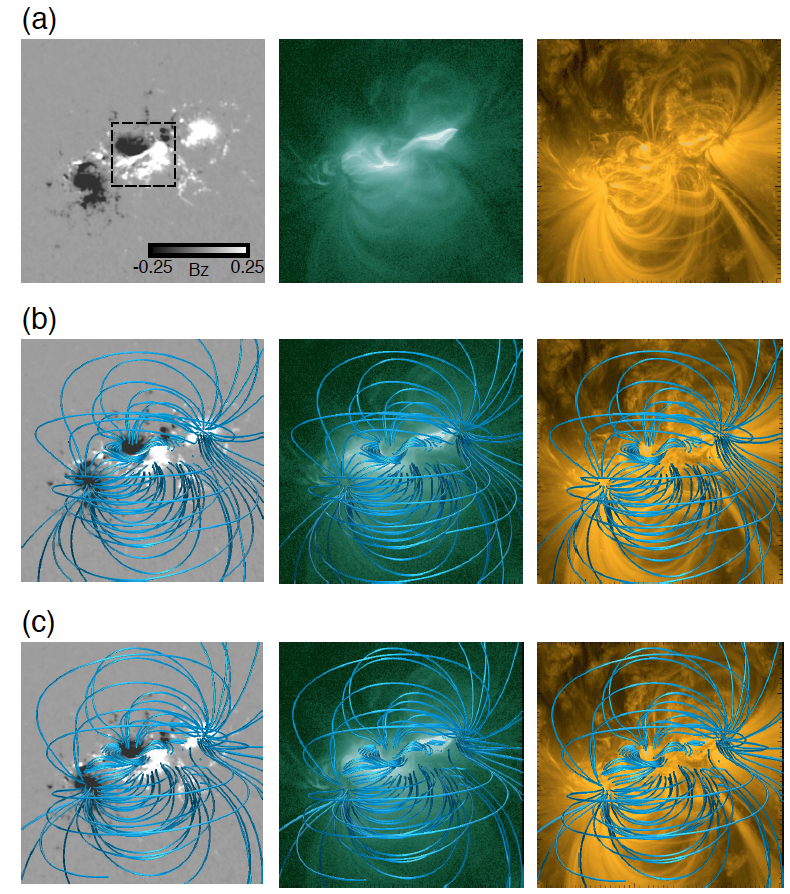}
  \caption{
           (a) Photospheric vector field (left), and EUV image with 94 {\AA}
               (middle), with 171 {\AA}(right) observed at 00:00 UT on 2011 
               February 15, taken by HMI and AIA on board {\it SDO}, are 
               shown, respectively. These sizes are in the range of 216 
               $\times$ 216 (Mm$^2$) and observed tangential components reside
               in the black dotted square 
               (79.2 $\leqq$ x $\leqq$ 136,8, 86.4 $\leqq$ x $\leqq$ 144)(Mm) 
               during the calculation through this paper except (c), while 
               other areas are fixed by the potential field. The value of 
               $B_z$ is normalized by 2500(G), non-dimensional value 0.25 
               corresponds to 625(G).
           (b) Field lines of the NLFFF are plotted over each image.
           (c) Field lines of the NLFFF which is reconstructed using whole 
               vector field, not partial reconstruction such as (b), are 
               plotted over each image.
           }
  \label{f1}
  \end{figure}

   We first show the 3D magnetic structure of NLFFF and compare it with EUV 
  images taken by AIA/{\it SDO}. We call this NLFFF calculation Run A and 
  all of Runs in this study are summarized in table \ref{tbl-1}. The upper 
  panels in Figure \ref{f1} (a) shows the vector field map obtained from HMI, 
  and EUV images with 94 {\AA} and 171 {\AA}, respectively. The 3D field lines 
  are plotted over each image shown in the middle panels in 
  Figure \ref{f1}(b). In the lower corona, we found that the strongly sheared 
  field lines are formed above PIL located between the sunspots in the central
  area. On the other hand, although the large loops extending to upper corona 
  can roughly capture the corona loops in AIA images, these reconstructed 
  field lines seem to be deviating from them as altitude increases. 

   We perform further NLFFF reconstruction on the basis of whole vector field, 
  not partial reconstruction such as in Figure \ref{f1}(b), in order to 
  further explore the reason of this problem. Figure \ref{f1}(c) shows the 
  NLFFF according the above manner, not showing the noticeable difference from
  that in Figure \ref{f1}(b). Consequently, the potential field and NLFFF do 
  not work effectively well on the reconstruction of the magnetic field in the
  upper corona. This is already suggested by \cite{2014ApJ...780..101I} that 
  it might be difficult to reconstruct these overlying field lines exactly 
  because they cannot keep steady state due to that they are expanding to the 
  further upper area pointed out by \cite{2011ApJ...731..122M}.
   
 \subsubsection{Physical Properties of the NLFFF}
\begin{figure}
  \epsscale{.95}
  \plotone{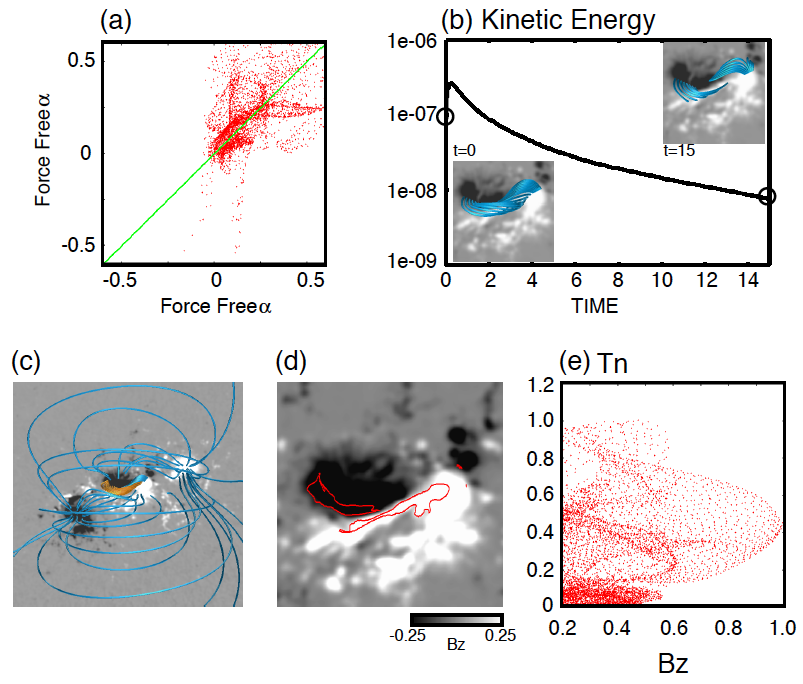}
  \caption{
          (a) Distribution of force-free $\alpha$ map of the NLFFF. The closed 
              field lines are focused and estimated in the central area 
              within the dashed square in left panel of Figure \ref{f1}(a). 
              Vertical and horizontal axes represent the values of the 
              force-free $\alpha$ in opposite footpoints on each field line. 
              These values are estimated in the plane at 720 km above the 
              photosphere, and the field lines are traced from the region in 
              which the values of the $B_z$ are less than $-$0.1 (=-250G). 
              Green line indicates the function of $y=x$.    
          (b) Temporal evolution of kinetic energy in Run B. Two insets 
              represent the 3D field lines at $t$=0, and $t$=15, respectively.
              Gray shows the $B_z$ distribution.
          (c) 3D field lines in the NLFFF. Orange lines have the twist value 
              more than half-turn ($T_n>0.5$) while the field lines with less 
              than half-turn twist ($T_n<$0.5) are plotted in blue.
          (d) The red line indicates the contour of half-turn twist
              ($T_n$=0.5). The inside of it is occupied by the strongly 
              twisted lines ($T_n >$0.5).
          (e) Twist values are plotted on each positive $B_z$ value more than 
              0.2(=500(G)).
          }
  \label{f2}
  \end{figure}

  Next, we show the physical properties of NLFFF, in particular, focusing on 
  how much the reconstructed field is in a force-free state, its stability, 
  and how much magnetic twist is accumulated in it. Figure \ref{f2}(a) shows 
  a distribution of the force-free alpha where the vertical and horizontal 
  axes represent the values of the force-free $\alpha$ estimated on opposite 
  footpoints of each field line. Like \cite{2013ApJ...770...79I}, these 
  force-free alpha are estimated at 720 km approximately {\it i.e.} 1 grid 
  size in this case above the photosphere, and the field lines are traced 
  from the region where $B_z<-$250 G. This result seems to be slightly larger 
  scatter than that in previous our result of 
  \cite{2013ApJ...770...79I}(see Figure \ref{f2}(a)) even though many points 
  are aligned close to the line of y=x plotted in green. This might be due to 
  the fine force-free $\alpha$ compared to previous one residing in the vector
  field reduced by 8$\times$8 binning process.
 
  We perform an MHD simulations using reconstructed NLFFF as an initial 
  condition to find a state in details of the NLFFF where the tangential 
  components of the magnetic field are released on the bottom boundary, 
  employing the constant resistivity. These calculation are named as Run B. 
  Figure \ref{f2}(b) shows a temporal evolution of the kinetic energy 
  $(\rho |\vec{v}|^2)/2)$ for Run B. Its value increases initially because 
  NLFFF cannot meet an equilibrium state exactly in addition to further 
  deviate from the initial state due to the release of the tangential 
  components of the magnetic field on the bottom surface, consequently the 
  velocity is generated. On the other hand, after that, this profile being 
  decrease as time goes on, which suggests that the reconstructed field is 
  being back to the potential field even on the disturbances derived from the 
  velocity. Two insets in Figure \ref{f2}(b) represent the 3D view of the 
  field lines at $t$=0 and $t$=15, marked by the circles. We clearly see that 
  the twisted field lines at $t$=15 become looser than the initial state 
  $t$=0, {\it i.e.,} the NLFFF is in stable state and never shows the drastic 
  dynamics as seen in the observations.

  We also investigate the magnetic twist that is here defined by
  \begin{equation}
   T_n =  \frac{1}{4\pi}\int \alpha dl,
  \end{equation}
  $\alpha$  and $dl$ are force-free 
  $\alpha$(= $\vec{J} \cdot \vec{B}/|\vec{B}|^2$) and a line element of a field 
  line, respectively. Figure \ref{f2}(c) shows the 3D field lines over the 
  $B_z$ distribution. Orange lines represent the strongly twisted lines over the
  half-turn twist which are traced inside of the red contours ($T_n=$0.5) shown 
  in Figure \ref{f2}(d). Figure \ref{f2}(e) represents the twist distribution 
  on the positive polarity where these twists are plotted on the intense $B_z$ 
  more than 0.2 (=500G). This result indicates that the most of twists have 
  a value less than one-turn, suggesting the stable against the kink mode 
  instability. These results are quite same as previous our result in 
  \cite{2013ApJ...770...79I} even though some parts of NLFFF model are 
  different.

% ------------------------------------------------------------------------------
 \subsection{Formation of the Strongly Twisted Lines in NLFFF}
% ------------------------------------------------------------------------------
   From above section showing the physical properties of NLFFF, we found that 
  the NLFFF never shows a dramatic dynamics or eruption away from the solar 
  surface, which is  contrary to the observation 
  (\citealt{2011ApJ...738..167S}). 
  Therefore, there might be mechanism that excites the stable NLFFF into dynamic 
  unstable phase to cause the observed eruptive phenomena. So we perform the MHD 
  relaxation process using the NLFFF as an initial condition, which is similar 
  manner with NLFFF calculation 
  in Run A except a velocity limit is released and an anomalous resistivity 
  described in equation (\ref{ano_res}) is set in this calculation, in order to 
  form a further strongly twisted lines more than one-turn twist. Because the 
  anomalous resistivity plays a role on enhancing the reconnection in the 
  strong current region, we can expect it to generate the more strongly twisted 
  lines through the reconnection than that formed in the NLFFF. Because the 
  magnetic field  is fixed at the bottom boundary during this calculation,  
  more strongly twisted lines are formed without changing the original 
  horizontal components on vector field, {\it i.e.,} map of the observed 
  force-free $\alpha$ on the vector field is not changed. We call this 
  calculation Run C. 

\begin{figure}
  \epsscale{.9}
  \plotone{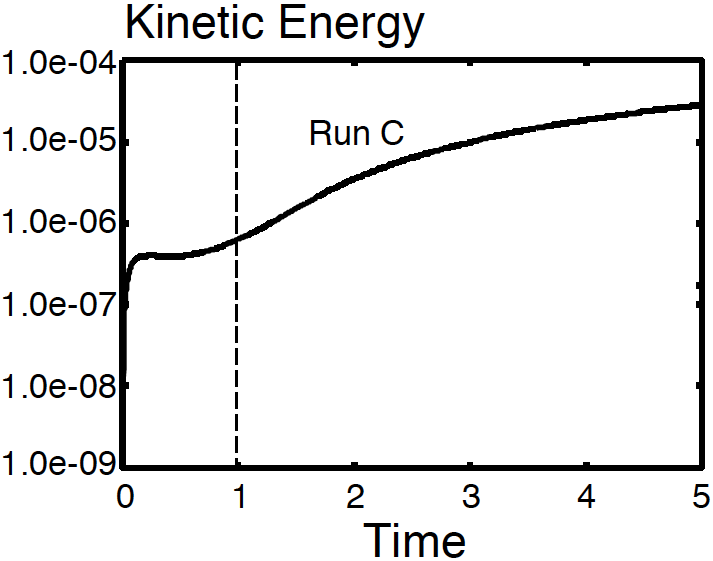}
  \caption{
          Temporal evolution of kinetic energy for Run C. The vertical dashed 
          line indicates $t$=1 at which the magnetic field is set up as an 
          initial condition in Run D.  
          }
  \label{f3}
  \end{figure}

  Figure \ref{f3} shows a temporal evolution of the kinetic energy for Run C.  
  We clearly see the kinetic energy starts to accelerate after $t$=0.5 and then 
  increasing as time goes on. The result in Run C indicates that solution 
  is deviating from the NLFFF, suggesting it converts into the dynamic phase 
  through the reconnection due to the anomalous resistivity. 

\begin{figure}
  \epsscale{.85}
  \plotone{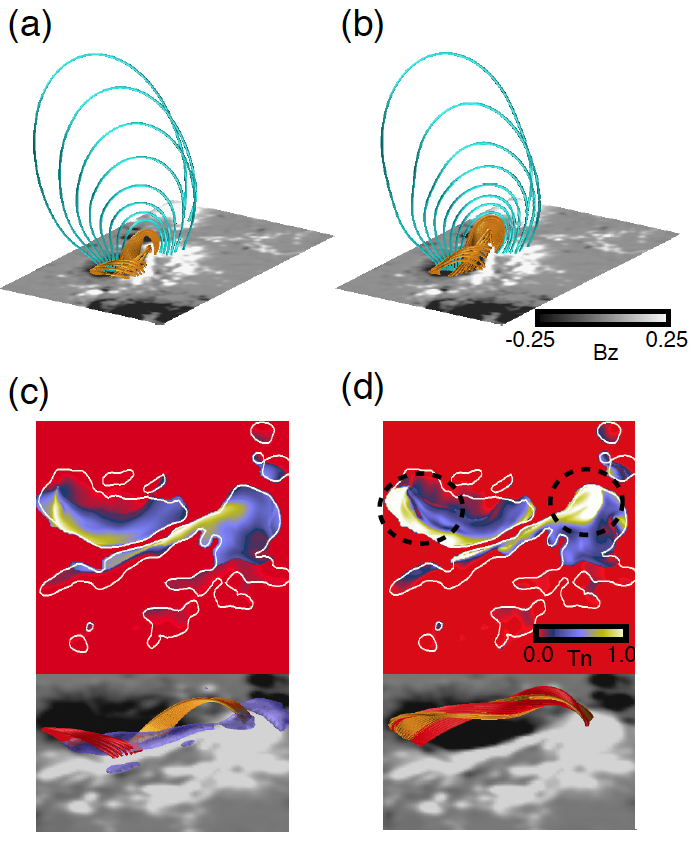}
  \caption{
         (a) and (b) show the 3D view of the field lines at $t$=0 and $t$=1, 
         respectively in Run C where these field lines are traced from same 
         positions. The orange lines represent the strongly twisted lines with 
         more than half-turn at $t$=1 while blue lines represent overlying 
         field lines surrounding the twisted lines. Upper panels in (c) and 
         (d) show the twist distributions mapped on the bottom surface at 
         $t$=0 and $t$=1, respectively where the white line is a contour of 
         $|B_z|$=0.25. The dashed circles indicate the location in which the 
         twist value is more than one-turn. Lower panels show the field lines 
         where red and orange field lines are traced from the regions 
         surrounded by the each dashed circle. Purple surface represents the 
         strong current $|\vec{J}|=$30 corresponding to a critical current 
         in the anomalous resistivity.   
          }
  \label{f4}
  \end{figure}

  In order to confirm this suggestion we note in previous section, we see 
  the 3D magnetic structure and estimate the twist of the magnetic field 
  lines at $t$=1 marked by the vertical dashed line in Figure \ref{f3}, 
  which just begins to convert into the dynamic phase. Figures \ref{f4}(a) 
  and (b) show the 3D view of the magnetic field at $t$=0 {\it i.e.,} NLFFF 
  and $t$=1. The twisted lines in orange at $t$=1 look changed slightly from 
  the initial NLFFF state even though the overlying field lines surrounding 
  these twisted lines almost keep their original configurations during this 
  period in Run C. We further see the distribution of the magnetic twist and 
  3D magnetic structure in Figures \ref{f4}(c) and (d) to understand the 
  more detailed magnetic structure at close to the PIL in the central area. 
  Figure \ref{f4}(c) shows that the most of twist values in NLFFF are less 
  than one-turn as discussed in the above section, on the other hand, the 
  strongly twisted lines with more than one-turn are generated at $t$=1 as 
  shown in Figure \ref{f4}(d), indicating the locations of their footpoints 
  marked by dashed circles in upper panel, and showing the field lines traced 
  from there in the lower panel. Interestingly, the locations of these 
  footpoints are close to those of sunquake detected by the helioseismology 
  (\citealt{2011ApJ...741L..35Z}), being triggered at the footpoint of the 
  erupting flux rope at early flare phase. In addition, the change of the 
  field lines topology from $t$=0 to $t$=1 allows us to be reminiscent of 
  the tether-cutting reconnection (\citealt{2012ApJ...745L...4L}, 
  \citealt{2012ApJ...745L..17W} and \citealt{2013ApJ...778L..36L}).  

  Therefore, this results might suggest that the tether-cutting reconnection 
  plays an important role on breaking the equilibrium sate before the flare.

% ------------------------------------------------------------------------------
  \subsection{MHD Simulation of Eruptive Magnetic Fields}
% ------------------------------------------------------------------------------
  \subsubsection{3D dynamics of Eruptive Magnetic Fields}
   We suggested in above section that the magnetic field with the strongly 
  twisted line over one-turn can break the equilibrium condition and make the  
  NLFFF convert into the dynamic phase. However its time evolution is not 
  consistent in that simulation because the three components of the magnetic 
  field are fixed on the boundaries, which is clearly over condition meaning 
  that the dynamics might be shown properly due to the inconsistent boundary 
  condition. In this section, we therefore perform the MHD simulation using 
  the magnetic field shown in Figure \ref{f4}(b), which is braking the 
  equilibrium state, as an initial condition where the tangential components 
  are released on all the boundaries to discuss its time evolution. This 
  calculation is called Run D.

\begin{figure}
  \epsscale{.9}
  \plotone{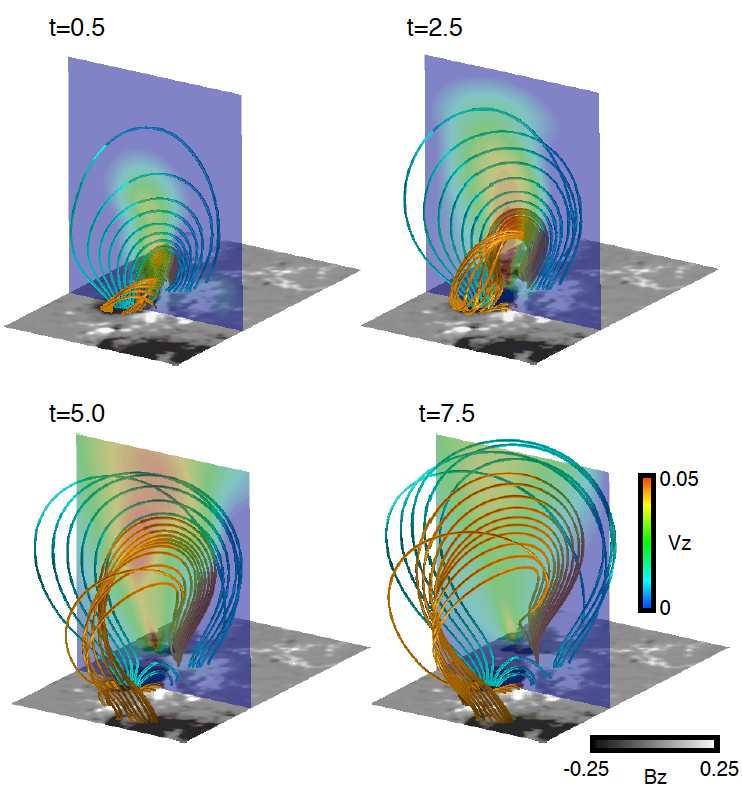}
  \caption{
           3D dynamics of the magnetic field lines. Orange lines represent 
           the twisted field lines with more than half-turn twist at $t$=0 
           in Run D, {\it i.e}, $t$=1 in Run C while blue lines represent 
           overlying field lines surrounding the twisted lines in orange. 
           $B_z$ distribution is drawn in gray and vertical velocity 
           distribution is mapped in color.
          }
  \label{f5}
  \end{figure}

    We first show an overview of the 3D Dynamics of the magnetic field lines
  for Run D with vertical velocity ($v_z$) in  Figure \ref{f5}. This result 
  shows that the twisted field lines with more than half-turn twist, being 
  formed at $t=$2 in Figure \ref{f4}(b), in orange erupt away from the solar 
  surface even though the tangential components of the magnetic field are 
  released on all the boundaries. Because of the release boundary condition, 
  the tangential components are deviating from that of original vector 
  field as time goes, which is relaxing toward them close to the potential 
  field. The NLFFF obtained from Runs B and C-1 never show these dynamics 
  with the constant resistivity, this result thus indicates that the 
  strongly twisted lines over the one-turn twist formed in NLFFF are 
  required in this active region to break the quasi-equilibrium state.

  \subsubsection{Comparison with Observational Data}
\begin{figure}
  \epsscale{1.}
  \plotone{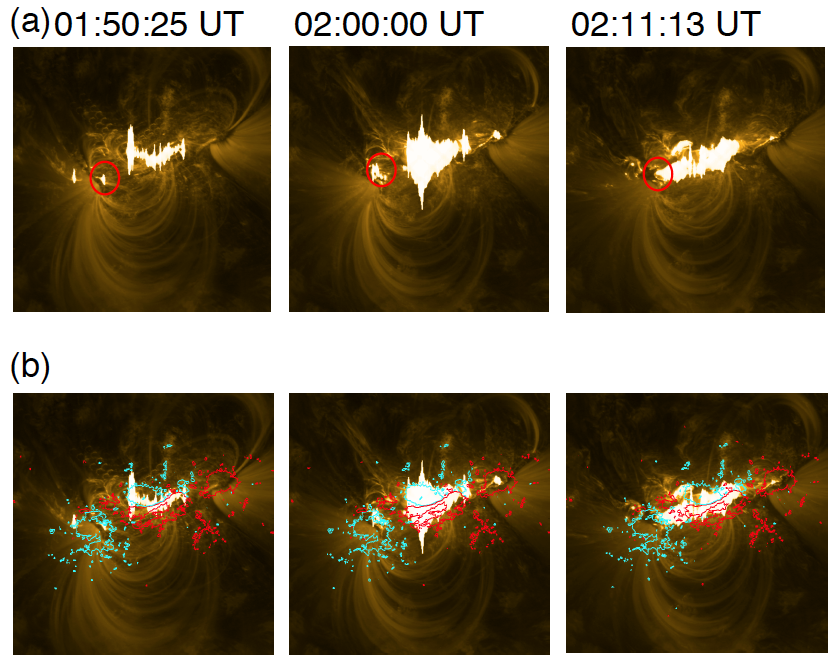}
  \caption{
          (a) EUV images with 171 {\AA} taken by AIA/{\it SDO} during the 
              X2.2-class flare, observed at 01:50:25 UT, 02:00:00 UT, and 
              02:11:13 UT, respectively on February 15. 
          (b) Red and blue lines are contours of $|B_z|$=0.25, plotted over 
              (a).
          }
  \label{f6}
  \end{figure}

    We compare these simulation results with observational data obtained 
  from to confirm a reliability of our simulation. Figure \ref{f6}(a) 
  shows the EUV images in 171 {\AA} taken by AIA/{\it SDO}. 
  Figure \ref{f6}(b) plots the contour of $|B_z|=0.25$ over the AIA 
  images. EUV is strongly enhanced at PIL between the sunspot located in 
  the central area, on the other hand, it is enhanced too at the location 
  marked by red circle, being away from the central area. This location 
  corresponds to edge of the negative polarity on the east side in 
  Figure \ref{f6}(b). From Figure \ref{f5}, we can see that the part of 
  one footpoints of the twisted field lines jump from the central area to 
  the edge of the negative polarity too after $t$=2.5 as suggested from 
  Figure \ref{f6}. Therefore, these observational results might support 
  our simulation.

\begin{figure}
  \epsscale{1.}
  \plotone{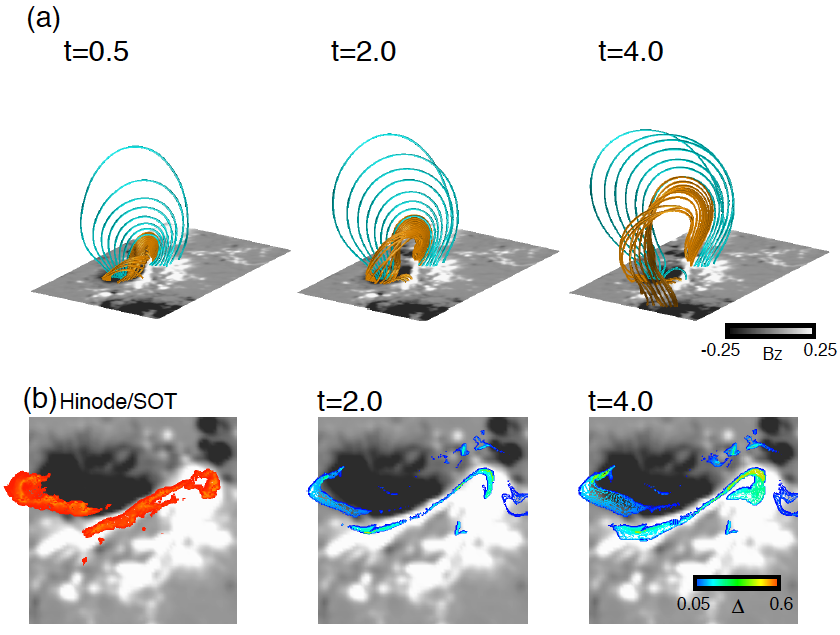}
  \caption{
           (a) Temporal evolution of the field lines from $t$=0.5 to $t$=4.0 
               in Run D, these are same format with Figure \ref{f5}. The gray 
               scale represents $B_z$ distribution.
           (b) Left:
               {\ion{Ca}{2}} image taken by FG/{\it Hinode} at 01:51 UT on 
               February 15 is plotted over the $B_z$ distribution.
               Middle and right:
               $\Delta$ Maps of spatial variance of the footpoint caused by 
               the reconnection of twisted line with more than one third 
               ($T_n$=0.3), which is defined in the equation (\ref{foot}), are 
               plotted at $t$=2.0 and $t$=4.0 over the $B_z$ distribution.
          }
  \label{f7}
  \end{figure}

   Next we compare our simulation results with {\ion{Ca}{2}} image taken by
  {\it Hinode} satellite. We focus on investigating the reconnection 
  of only the twisted lines with more than one third ($T_n$=0.3) because  
  \cite{2011ApJ...738..161I}, \cite{2012ApJ...760...17I}  and 
  \cite{2013ApJ...770...79I} suggested that these twisted line at close to 
  the PIL are related to the two-ribbon flares where the {\ion{Ca}{2}} image 
  is strongly enhanced. In addition to this, in this study we analyze it at 
  early eruptive phase as shown in Figure \ref{f7}(a) because in the late 
  phase the tangential components of the magnetic field on the bottom 
  boundary are deviating gradually from observational one due to the released 
  boundary condition. 

   Following \cite{2013ApJ...773..128T}, the reconnected field lines are 
  estimated using the spatial variance of the field lines connectivity, which 
  is allowed only by the magnetic reconnection. The formulation is defined by 
  the following equation;
   \[
   \delta(\vec{x}_0,t_n)  = |\vec{x}_{1}(\vec{x}_0, t_{n+1})
                                            - \vec{x}_{1}(\vec{x}_0, t_{n})| 
                            \quad (\mbox{$|T_n| \geqq 0.3$}),
  \]
  where $\vec{x}_{1}(\vec{x}_0, t_{n})$ is a location of one footpoint of
  each field line at time ${\it t_n}$, which is traced from another footpoint
  at $\vec{x}_0$. Eventually, we calculate
  \begin{equation}
    \Delta(\vec{x}_0,t) = \int_{0}^{t} \delta(\vec{x}_0,t_n)dt_n,
    \label{foot}
  \end{equation}
  where the enhanced region in $I(\vec{x}_0, t)$ indicates memories
  in which the reconnection took place dramatically in the twisted lines. 
  The left panel in Figure \ref{f7}(b) exhibits the {\ion{Ca}{2}} image taken
  by {\it Hinode}/SOT at the early flare phase and the middle and right panels 
  show the distribution of $I(\vec{x}_0)$ at $t$=2.0 and $t$=4.0. 
  Interestingly, the distribution of $I(\vec{x}_0)$ is similar to that 
  of the observed two-ribbon flares seen in the left panel in 
  Figure \ref{f7}(b). This result clearly shows that the two-ribbon 
  flare is enhanced due to the reconnection of twisted lines, which is 
  consistent with classical flare models(see \citealt{2011LRSP....8....6S}).  

\begin{figure}
  \epsscale{.7}
  \plotone{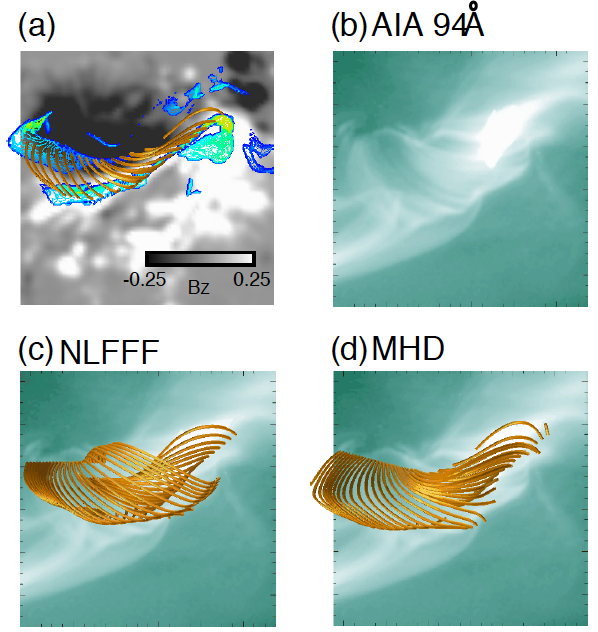}
  \caption{
          (a) Field lines are plotted with map on spatial variance of the 
              footpoint caused by the reconnection at $t$=5.0 being same 
              format with Figure \ref{f7}(b), over the $B_z$ distribution in 
              gray.
          (b) AIA image in 94 {\AA} taken by {\it SDO} observed at 02:29:50 
              UT on February 15.
          (C) Field lines of NLFFF, which are reconstructed from vector field 
              observed at 03:00 UT on February 15, are plotted over the AIA 
              image in (b).
          (d) Field lines from MHD simulation at $t$=10 are plotted over the 
              AIA image.  
          }
  \label{f8}
  \end{figure}

   Finally, we show the field lines structure after the flare. 
  Figure \ref{f8}(a) shows the field lines over the flare-ribbons obtained 
  from MHD simulation from which all of field lines are traced. Our simulation 
  shows the post flare loops on the flare ribbons as often seen in observed 
  EUV or {\ion{Ca}{2}} images. We compare the simulation results with the EUV 
  image with 94 {\AA} taken by AIA/{\it SDO} shown in Figure \ref{f8}(b).
  Before showing the comparison with the result of MHD 
  simulation, Figure \ref{f8}(c) shows the NLFFF after the flare, which is 
  reconstructed from vector field at 03:00 UT on February 15, plotted over 
  the EUV image. The NLFFF almost capture the post flare loops in EUV image
  shown in Figure \ref{f8}(b). On the other hand, the field lines structure 
  at $t$=10   obtained from MHD simulation slightly deviates from the 
  observed image and looks relaxation compared to those of NLFFF. Because the 
  tangential components of the magnetic field are released on the bottom 
  boundary, and then being toward close to the potential field. Nevertheless, 
  they roughly capturer the EUV image and this result would also support the 
  reliability of our simulation.
     
% ===============================================================================
  \section{Discussion}
% ===============================================================================
  \subsection{Formation of the CME}
\begin{figure}
  \epsscale{1.}
  \plotone{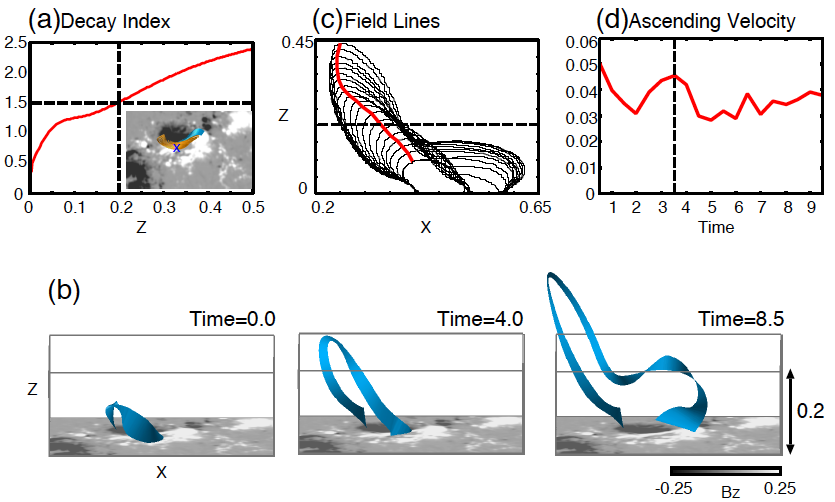}
  \caption{
           (a) Height profiles of the decay index on potential field are
               plotted, measured a cross in the inset in which the
               the NLFFF has strongly twisted lines (more than half-turn).
               The length is normalized by the 216(Mm), so h=0.2 corresponds 
               to 63.2(Mm).
           (b) A temporal evolution of selected twisted field lines are 
               plotted in blue ribbon. The top of box corresponds to a 
               critical height of the torus instability.
           (c) The same field lines in (b) are projected in x-z plane from 
               $t$=0.5 to $t$=10. The red line traces the loop top and dashed 
               line indicates the threshold height where twisted loop becomes 
               torus instability.
           (d) The time variance of the loop top, {\it i.e.,} the ascending 
               velocity of the twisted loop on appearance. 
          }
  \label{f9}
  \end{figure}

   Above section, the MHD simulation showed the twisted lines initially embedded
  in the lower corona successfully erupt from the solar surface. However, its 
  still unclear that these twisted lines grow up to the CME or not. In order to 
  clarify them, we discuss about the decay index profile being related to the 
  torus instability (\citealt{2006PhRvL..96y5002K}), which is an important 
  factor in determining whether the flux tube can escape from the lower corona 
  or not. The decay index refers to an estimation of a criterion for the torus 
  instability and is given by the following equation:  
  \begin{equation}
   n(z) = - \frac{z}{|\vec{B}|}\frac{\partial |\vec{B}|}{\partial z},  
  \end{equation}
  where the threshold is at approximately $n_c$=1.5 
  (\citealt{2007AN....328..743T}). Because it is difficult to separate  an 
  external field from NLFFF,  following \cite{2007ApJ...668.1232F} 
  or \cite{2010ApJ...708..314A}, the decay index on the potential field 
  is plotted in Figure \ref{f9}(a), marked by a cross in the inset, where 
  the strong twisted lines of NLFFF are embedded. From this result, 
  the threshold is located at approximately $h$ = 0.2 corresponding to 
  63.2(Mm). Figure \ref{f9}(b) shows a time evolution of selected twisted 
  field line with more than half-turn where at $t$=0.0 the twisted line 
  exists in the lower area corresponding to less than $h$ = 0.2 at the top 
  of the box. This indicates that the twisted line is stable against the 
  torus instability. On the other hand, as time goes on, these cross over 
  the critical height as a result of the initial eruption. This result might 
  give an important scenario related to the dynamics of the flare-CME.  

   We present more detailed temporal evolution of the same selected twisted 
  line in Figure \ref{f9}(c) projecting them at each time on 2D plane with 
  red line tracing a temporal evolution of the loop top. The dashed line 
  indicates the critical height of the torus instability. Figure \ref{f9}(d) 
  shows the time variation of the loop top {\it i.e.,} the ascending velocity 
  on appearance with the vertical dashed line where the twisted line is 
  passing the critical height. This result does not exhibit sudden 
  acceleration of the twisted line in this study, as shown in 
  \cite{2010ApJ...708..314A} and \cite{2010ApJ...719..728F}, even though the 
  loop top of the twisted line exceeds the critical height and its velocity 
  looks to being increase slowly after $t$=5.0.

   We might point out two causes to bring this problem. One is that the whole 
  box is not large enough to trace the twisted lines for a long time. In 
  addition, we set the rigid wall at all the boundaries, which is not able to 
  expel the overlying field lines to the outside of the numerical domain, 
  consequently the magnetic pressure is accumulated close to the boundary due 
  to the twisted lines keep pushing the overlying field lines. Another is due 
  to the bottom boundary condition {\it i.e.,} due to the release of the 
  tangential components. Torus instability is driven by the poloidal field of 
  the flux tube generated by the toroidal current carrying in it. However, the
  release condition for the tangential components relax the twisted lines, 
  which means that hoop force driving the twisted lines upward is loosing 
  gradually. These two factors preventing the twisted lines from the 
  acceleration might work effectively on the dynamics of them. In other word, 
  these twisted lines might have a potential for the solar eruption if 
  settling these problems on the size of numerical box and the boundary 
  condition. Therefore, as prospective work, it is important matter how to 
  fit the three component of the magnetic field obtained from the vector 
  field data consistently into an induction equation, {\it i.e.,} would be 
  required a data-driven simulation as presented in 
  \cite{2012ApJ...757..147C}.   
  
  \subsection{Depending on the Density Profile}
\begin{figure}
  \epsscale{.95}
  \plotone{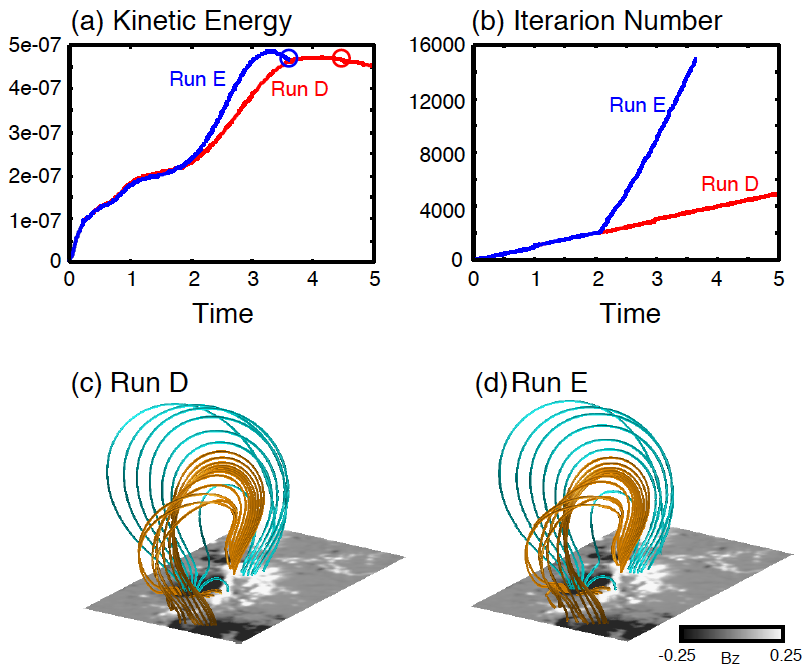}
  \caption{
           A temporal evolution of (a) kinetic energy and (b) iteration 
           numbers for Run D and Run E in red and blue, respectively.
           3D field lines structure in (c) Run D and (D) Run E are plotted.
           These formats are same as Figure \ref{f5}.
          }
  \label{f10}
  \end{figure}

    We check the dynamics of depending on the density formula by replacing the 
  equation (\ref{eq_den_NLFFF}) by the equation (\ref{eq_den_MHD}) where $\xi$ 
  is given by $1.0 \times 10^{-4}$. This calculation is called Run E.
  Figure \ref{f10}(a) shows a temporal evolution of the kinetic energy for 
  Run D and Run E, respectively. Both of them are a lot of similar profiles 
  except we can see a little bit deviation at the late phase. On the other 
  hand, Figure \ref{f10}(b) shows an temporal evolution of the iteration 
  number, which shows that Run E requires much more calculation time than that
  in Run D. 3D field lines structure marked by each circle in 
  Figure \ref{f10}(a) are shown in Figures \ref{f10}(c) and (d). As a 
  consequence, we conclude that the dynamics is almost same between them but 
  the cost performance of Run D is much better than Run E, as shown in 
  \cite{2006ApJ...645..742I}. This conclusion is also agreement with previous 
  works done by \cite{1996ApJ...466L..39A} and series of their papers.

  \subsection{Depending on the Resistivity Formula} 
\begin{figure}
  \epsscale{.95}
  \plotone{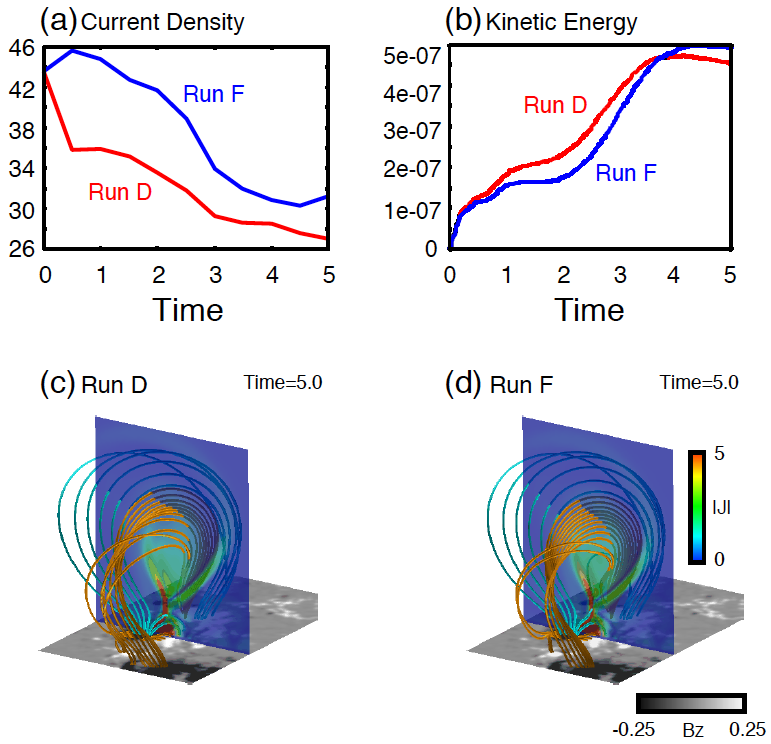}
  \caption{
           A temporal evolution of (a) maximum current density 
           ($|\vec{J}|_{max}$) measured above 3600 (km) corresponds to 5 grid 
           above the photosphere and  (b) kinetic energy for Run D and Run F in 
           red and blue, respectively. The 3D field line structure with 2D 
           $|\vec{J}|$ map at $t$=5.0 are plotted in (c) Run D and (d) Run F. 
           Field lines format is same as Figure \ref{f5}.  
          }
  \label{f11}
  \end{figure}

   We further perform an MHD calculation (Run F) replacing the anomalous 
  resistivity by constant one to show a dynamics of depending on the 
  resistivity formula too. Figure \ref{f11}(a) shows a temporal evolution 
  of the maximum value of current density (=$|\vec{J}|_{max}$) measured 
  above 3600(km) for Run D and Run F, respectively, which can exclude the 
  strong current density existing close to the photosphere. Although the 
  value of Run D initially undergoes sudden decrease and is keeping less 
  than one in Run F during  the calculation, this sudden decrease is 
  probably due to the diffusive effect derived from the anomalous 
  resistivity. On the other hand, because the critical current set in the 
  anomalous resistivity corresponds to the value 30, it works around by 
  $t$=3. Nevertheless, temporal evolutions of the kinetic energy for Run D 
  and Run F shown in Figure \ref{f10}(b) are quite similar profiles 
  although we can see a little bit deviation between them. Furthermore 
  3D field lines structure and color map of $|\vec{J}|$ for Run D and 
  Run E shown in Figures \ref{f10}(c) and (d) are almost same too even at 
  $t$=5.0. Form these results, in this case the anomalous resistivity dose 
  not work effectively well on the dynamics while the twisted lines are 
  ascending. However, note that the reconnection is very important to form 
  the strongly twisted lines in the lower corona to launch from the solar 
  surface, after that it would take place only accompanied with ascending 
  of the twisted lines pointed out by \cite{2006ApJ...645..742I}.

% =============================================================================
% Summary
% =============================================================================
  \section{Summary}
   We performed the MHD simulation combined with NLFFF and show an overview of 
  the dynamics of the magnetic field at early eruptive phase of the X2.2-class 
  flare, and compared with observational results. The NLFFF is in a stable 
  state against the kink mode instability (also confirmed by 
  \citealt{2013ApJ...770...79I}) and even any other perturbations as long as 
  reconnection dose not occur in the NLFFF.  On the other hand, when the 
  tether-cutting reconnection in NLFFF could form the strongly twisted lines 
  with more than one-turn twist, then we found that it plays an important role
  on breaking the quasi-steady state in NLFFF. Consequently the NLFFF converts 
  into the dynamic stage and the twisted lines can erupt away from the solar 
  surface. This scenario is already suggested by the previous studies on 
  analysis of NLFFF in AR 11158 by  
  \cite{2012ApJ...745L..17W} ,  
  \cite{2012ApJ...745L...4L},  
  \cite{2013ApJ...770...79I}  and 
  \cite{2013ApJ...778L..36L}. 
   As a result for comparing with the observations, distribution of the 
  observed two-ribbon flares are explained well by map on the spatial variance 
  of the footpoints due to the reconnection of the twisted lines. In addition 
  to this, the post flare loops seen in the AIA image are captured by the 
  field lines after the reconnection on the MHD simulation. Thus, these 
  results could support the reliability of our simulation results. 
         
   However, the some detailed dynamics on the twisted lines have not yet been 
  revealed. One is that how much dose the flux tube need the magnetic twist 
  just before launching from the solar surface, in order to reach the 
  threshold height of the torus instability. In this study, although only one 
  case (Run D) was performed and then presented here, we have to investigate 
  the dynamics of another twisted lines having the different twist values 
  initially, which is extended from  this study. Then we can expect to find 
  the critical twist leading the torus instability. Another is to reveal the 
  complicated magnetic reconnection while the twisted line is ascending. The 
  detailed analysis extended from this paper will answer this question 
  definitely. 

   On the other hand, our simulation left some problems. Although the 
  twisted lines cross over the threshold height of the torus instability, 
  we could not see the sudden acceleration shown in the previous theoretical 
  studies done by {\it e.g.}, \cite{2007AN....328..743T}, 
  \cite{2010ApJ...708..314A}, and \cite{2010ApJ...719..728F}. 
  In order to answer this problem, we should extend size of the numerical 
  domain, modify or develop an advanced boundary condition fitting the three 
  component of the magnetic field on the photospheric magnetic field 
  consistently into the induction equation , such as a data-driven simulation 
  would be required (\citealt{2012ApJ...757..147C}).
     
   In addition, in this study the triggering reconnection to produce the 
  twisted flux tube is induced by the anomalous resistivity. However, in 
  general, photospheric motion or new emerging flux drives pre-existing 
  strongly twisted lines into the dynamic phase ({\it e.g.}, 
  \citealt{1989ApJ...343..971V}, \citealt{1995JGR...100.3355F}). Therefore, 
  in the future, we must develop a simulation considering  this observational 
  information. Doing so will allow further physical insight into the onset 
  and dynamics of flare-CME. 
                                           
% ========================================================================
% Acknowledgements
% ========================================================================
  \acknowledgments
   We are grateful to anonymous referees for helping us improve and polish 
   this paper. S.\ I.\ was supported by the International Scholarship of Kyung 
   Hee University. This work was supported by the Korea Astronomy and Space 
   Science Institute under the R $\&$ D program (project No.2013-1-600-01) 
   supervised by the Ministry of Science, ICT and Future Planning of the 
   Republic of Korea. G.\ S.\ C. was supported by the National Research 
   Foundation grant NRF-2010-0025403. The computational work was carried 
   out within the computational joint research program at the Solar-Terrestrial 
   Environment Laboratory, Nagoya University. Computer simulation was performed 
   on the Fujitsu PRIMERGY CX250 system of the Information Technology Center, 
   Nagoya  University. Data analysis and visualization are performed using 
   resource of the OneSpaceNet in the NICT Science Cloud. We are sincerely 
   grateful to NASA/SDO and the HMI and AIA science team. Hinode is a Japanese 
   mission developed and launched by ISAS/JAXA, with NAOJ as domestic partner 
   and NASA and STFC (UK) as international partners. It is operated by these 
   agencies in co-operation with ESA an NSC (Norway).

\clearpage

% -------------------------------------------------------------------------
% Exact & Potential Field for Low & Lou Solution
% -------------------------------------------------------------------------
  \clearpage

% -------------------------------------------------------------------------
% Temporal Evolution of Kinetic Energy. Twist 
% -------------------------------------------------------------------------
  \clearpage
  
% -------------------------------------------------------------------------
% 3D dynamics of field lines
% -------------------------------------------------------------------------
  \clearpage

% -------------------------------------------------------------------------
% Flare-Ribbon & Reconnection
% -------------------------------------------------------------------------
  \clearpage

% -------------------------------------------------------------------------
% 3D Field Lines & Velocity
% -------------------------------------------------------------------------
  \clearpage
%
% -----------------------------------------------------------------------------
% AIA
% -----------------------------------------------------------------------------
  \clearpage

% ---------------------------------------------------------------------------- 
% Numerical Flare Ribbons  
% ----------------------------------------------------------------------------
  \clearpage

% ----------------------------------------------------------------------------
% Comparing with NLFFF after the Flare and Flare loops  
% ----------------------------------------------------------------------------
  \clearpage

% ---------------------------------------------------------------------------- 
% Possibility of a Solar Eruption due to Trous instability 
% ---------------------------------------------------------------------------- 
  \clearpage

% ----------------------------------------------------------------------------  
% Depending on the density                      
% ----------------------------------------------------------------------------  
  \clearpage

% ---------------------------------------------------------------------------- 
% Depending on a resistivity 
% ---------------------------------------------------------------------------- 
  \clearpage

% -----------------------------------------------------------------------------
% Equation and Boundary Conditions...for MHD Vs. NLFFF
% -----------------------------------------------------------------------------
  \begin{table}
  \begin{center}
  \caption{Each Run for NLFFF or MHD simulation, employing equations of density 
           evolution, resistivity formula, initial and boundary conditions. 
           \label{tbl-1}}
  \begin{tabular}{crrrrr}
  \tableline\tableline
  Run & type & density & resistivity & Initial condition & Boundary condition \\
  \tableline
  Run A  & NLFFF          & eq.(\ref{eq_den_NLFFF}) & eq.(\ref{eta_nlfff})                & Potential field  & Fix \\
  Run B  & MHD Simulation & eq.(\ref{eq_den_NLFFF}) & constant  & NLFFF & Release \\
  Run C  & MHD Relaxation & eq.(\ref{eq_den_NLFFF}) & anomalous & NLFFF & Fix \\
  Run D  & MHD Simulation & eq.(\ref{eq_den_NLFFF}) & anomalous & $t=$1 in Run C          & Release \\
  Run E  & MHD Simulation & eq.(\ref{eq_den_MHD})   & anomalous & $t=$1 in Run C              & Release \\
  Run F  & MHD Simulation & eq.(\ref{eq_den_NLFFF}) & constant  & $t=$1 in Run C              & Release \\
  \tableline
  \end{tabular}
  \end{center}
  \end{table}

\end{document}